\documentclass[sigconf,nonacm]{acmart}
\AtBeginDocument{%
  \providecommand\BibTeX{{%
    \normalfont B\kern-0.5em{\scshape i\kern-0.25em b}\kern-0.8em\TeX}}}

\setcopyright{none}

\begin{document}

\title{Fuzzified advanced robust hashes for identification of digital and physical objects}

\author{Shashank Tripathi}
\affiliation{%
  \institution{HAW Hamburg}
  \city{Hamburg}
  \country{Germany}}
\email{shashank.tripathi@haw-hamburg.de}

\author{Volker Skwarek}
\affiliation{%
  \institution{HAW Hamburg}
  \city{Hamburg}
  \country{Germany}}
\email{volker.skwarek@haw-hamburg.de}

\renewcommand{\shortauthors}{Shashank and Skwarek, et al.}

\begin{abstract}
    \textbf{
     With the rising numbers for IoT objects, it is becoming easier to penetrate counterfeit objects into the mainstream market by adversaries. Such infiltration of bogus products can be addressed with third-party-verifiable identification. Generally, state-of-the-art identification schemes do not guarantee that an identifier e.~g.~ bar codes or RFID itself cannot be forged. This paper introduces identification patterns representing the objects intrinsic identity by robust hashes and not only by generated identification patterns. Inspired by these two notions, a collection of uniquely identifiable attributes called quasi-identifiers (QI) can be used to identify an object. Since all attributes do not contribute equally towards an object’s identity, each QI has a different contribution towards the identifier. A robust hash developed utilising the QI has been named \emph{fuzzified robust hashes (FaR hashes)}, which can be used as an object identifier. Although the FaR hash is a single hash string, selected bits change in response to the modification of QI. On the other hand, other QIs in the object are more important for the object’s identity. If these QIs change, the complete FaR hash is going to change. The calculation of FaR hash using attributes should allow third parties to generate the identifier and compare it with the current one to verify the genuineness of the object.
    }

\end{abstract}


\maketitle

\section{Introduction}

Since the last decade, there has been a significant surge in IoT devices or objects. The number is expected to grow beyond 25.4 billion devices by 2030 \footnote{https://www.statista.com/statistics/1183457/iot-connected-devices-worldwide/}. Manufacturers are producing numerous objects to keep up with the market demand. However, high security and attack prevention are desired to reduce the threats to IoT devices. A study\footnote{\url{https://www.itpro.co.uk/security/22804/hp-70-of-internet-of-things-devices-vulnerable-to-attack}} by the hardware and software company HP in 2014 has revealed that 70\% of IoT devices are vulnerable to attacks. With the rising IoT market, the volume of attacks is expected to rise. There could be different targets for attacking an IoT; namely, attacking the device, network or server where data is stored. To conduct any of these attacks, the adversary should first install or download a malicious file at the target location. There are several methods to detect this action and prevent it. One such way is to scan all the memory locations in the concerned system. It is a common practice to compare the content of all the files in the system with the database records of malicious files that have been used in past attacks. For efficiency, different kinds of hashes are generated and stored in the database. During the scan, the hashes of each file are compared to the recorded hashes. Over time, the attackers have learnt to bypass this measure by changing a few bits in the malicious file. Due to this change, pure cryptographic hashes like SHA and MD5 change entirely. While hashes like ssdeep \cite{ssdeep} calculate the hashes for different blocks in a file. The block size is calculated based on the content of that file. 

This paper proposes a hash inspired by the connotation of \emph{identity}. The idea is that each physical object or digital object possesses attributes that together make them unique. 
Generally, identity is a representation of an entity in a specific domain. In other words, identity is a combination \cite{Jsang2007UsabilityAP} of identifiers and credentials. Name, birthday, food habits, honesty, and temperament for a human, whereas, for a digital object, the identity is defined by its interaction with humans and other objects. An object can be uniquely addressed by its quantifiable and unquantifiable attributes. Additionally, such identity-based hash can uniquely represent an IoT device for enhancing security against counterfeit objects.

 \begin{figure}
     \centering
     \includegraphics[scale=0.8]{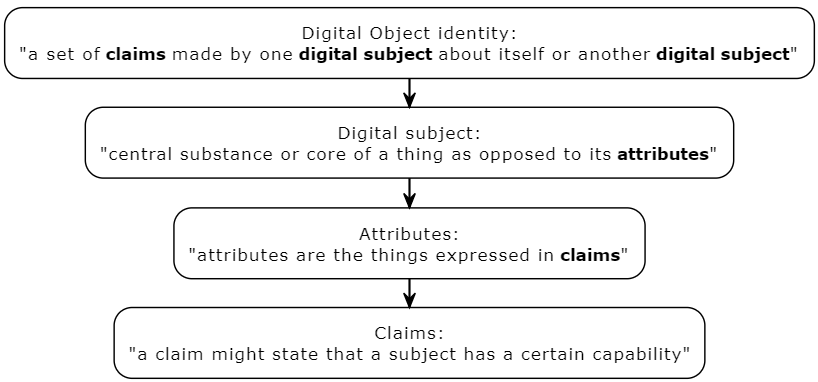}
     \caption{Digital object identity definitions quoted from \cite{cameron2005laws} }
     \label{fig:object_identity_definitions}
 \end{figure}

While explaining the identities for digital objects, the author \cite{cameron2005laws} has specified object identities, summarised in Figure \ref{fig:object_identity_definitions}. The diagram concludes with terms associated with an object's identity, namely, claims and attributes. An attribute is a piece of an entity's information regarding its quality or property. A combination of multiple attributes can uniquely identify most entities in the real world. Such attributes are called quasi-identifier (QI) \cite{dalenius1986finding.quasi-identifier}.

\begin{figure}
    \centering
    \includegraphics[scale=0.19]{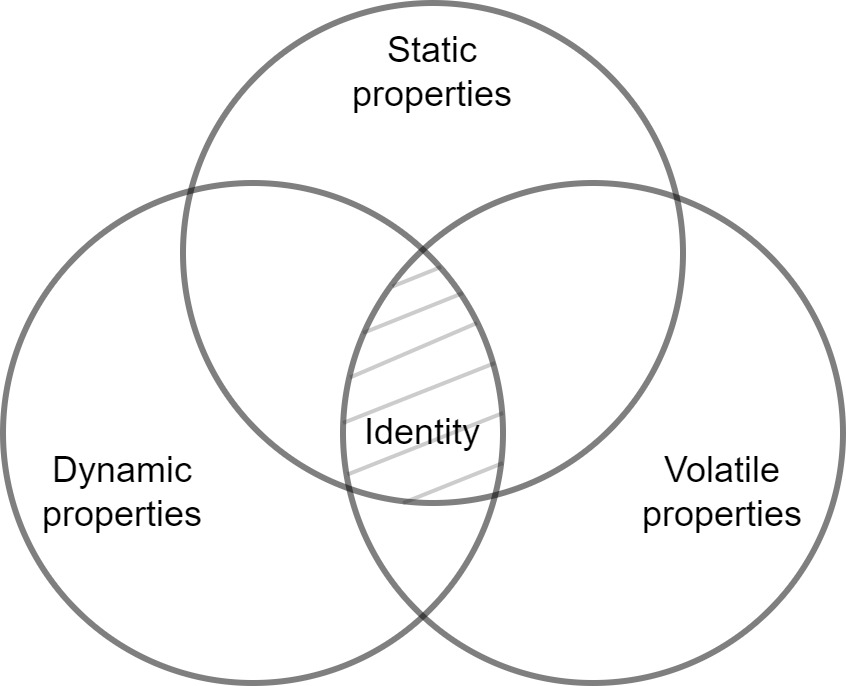}
    \caption{Symbolic Venn-diagram representation of the identity as a mixture of different attributes of an object}
    \label{fig:venn_diagram}
\end{figure}

In this paper, the object attributes (or QIs) have been divided into three categories: static, dynamic and volatile. Static QIs are the strictly required properties of an object. A modification in these attributes changes the identity of objects to new identites. Secondly, dynamic QI define a change of identity attributes within a defined range - such as the visual appearance of a face with by aging. It alters an object's identity only insignificantly - or strictly spoken: the overall identity stays the same. Lastly, volatile QIs are short-term attributes or set of characteristics. They do not affect the overall identity of an object as single item but only an as significant change over the whole set of volatily attributes. As depicted in Figure \ref{fig:venn_diagram}, the common intersection of the three attributes is the area of interest for this research. Classifying the attributes of a particular object in these categories is beyond this research's scope, but in this paper, different examples will be taken to demonstrate the identity concept.

Identity management refers to the processes and technologies dealing with authentication and authorisation. The motivation behind the research is an attempt to address the requirements of identity management systems (IDMS). The main requirements of IDMS classified by the literature \cite{torres2012identity} are: privacy, security, usability, mobility, affordability, trustworthiness, law enforcement, and interoperability. The requirements are further discussed in the following sections. The identity scheme designed in the current research considers these requirements and proposes an identifier model that addresses most of them.

The identifiers are generally alphanumeric in state-of-the-art identifiers like RFID (Radio Frequency Identification), Barcode/QR-code, IPv4/v6, electronic product code (EPC), and machine learning methods. Such identifiers are often uncorrelated with the object’s identity. Other identification methods like document fingerprinting, robust hashing and git’s naming convention take into account the document’s content to create a hash-based identifier.

A hash is a fixed-length string that uniquely identifies the input data. The hash produces a string of the same length irrespective of the input data length. It is expected that the hashes resulting from different sources should not collide easily. Although, an absolute collision-free hashing algorithm is impossible, as argued with the birthday paradox \cite{flajolet1992birthday}. Even good hashing algorithms are a function of the contemporary computation capabilities of a given generation and cannot remain secure forever.
Nevertheless, an excellent collision-resistant algorithm can guarantee that finding a collision is computationally infeasible. Thus, it is tough to find two same inputs for a given hash. MD (Message-Digest algorithm) and SHA (Secure Hash Algorithm) are typical examples of established families of hashing algorithms that keep ruling out new versions with time. The standard hashing and hash-based identification algorithms have been briefly discussed in the section \ref{sec:identifier_creation}. Since this paper discusses multiple kinds of hashes, the hash described in this paragraph will be called the standard hash.

Robust hashing \cite{venkatesan2000robust} exhibits different behaviour compared to the standard hashes. The standard hashes are designed to produce completely different outputs for two marginally dissimilar inputs. Whereas, in the case of robust hashes, two visually similar images with little contrasts produce the same hashes. It was achieved by dividing the image using tiles of random sizes and then calculating the hash of a group of tiles. The small changes, negligible enough not to get reflected in the tiling process, would produce the same encoded image and thus produce the same hash. 

This paper extends the idea of robust hashes by introducing fuzzified advanced robust hashes (FAR hash). Thereby, two main features are introduced. Firstly, the FAR hashes are partly modifiable. Secondly, it can behave like a standard hash, meaning the hash string changes entirely. These features can be used to build an identity model that answers the following research question:

\begin{enumerate}
    \item [\textbf{RQ1}] \textbf{How to uniquely and dynamically identify digital and physical objects based on their attributes?}
\end{enumerate}

 \section{Identity and Identification concept}
\label{sec:identity}

Throughout lifetime, different settings and experiences develop the identity of an entity. An identity consists of properties in virtue of which it is that entity. Implying that if these properties are modified, they will cease to be that entity and become something else (c.f. page 12, \cite{JamesD.Fearon.identity}). The “Ship of Theseus” paradox also raises a similar question. The paradox states whether a ship with all its components replaced remains the same ship \cite{Wortmann.2016.theseus.paradox}. It should be noted that these properties do not necessarily need to be publicly visible or even perceivable: they could also be experience based or hidden inside the entity and can only by perceived as an overall reaction of a trigger. For instance, if a car owner gets the car’s engine replaced without modifying the car’s optics, the car remains visually indistinguishable from the other cars of the same model while showcasing different interactivity. Is the altered version of the car still identifiable based on the original specifications? Or the rather philosophic question "Who or what are you?"

Depending on the production environment and usage patterns, every car is unalike at the built time and during application. Therefore, every car of the same model is differentiable from the other cars. The same is true for other objects as well. However, this paper focuses only on object identities. Additionally, this paper does not try to justify each object’s uniqueness but demonstrates how to quantify identities given the object’s uniqueness.

\begin{figure}
    \centering
    \includegraphics[scale=0.4]{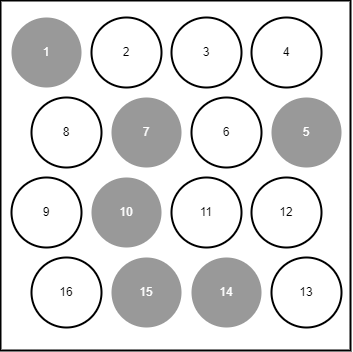}
    \caption{An object as a combination of multiple attributes. The circles represent different attributes. The shaded circles represent attributes that can serve as a QI.}
    \label{fig:object_as_mix_of_attributes_drawio}
\end{figure}

As established, object identity can be defined by combining unique attributes or QI. The QI are the attributes of an object that might not be unique in isolation but, together with other QIs, uniquely identify an object. 
 
In Figure \ref{fig:object_as_mix_of_attributes_drawio}, the square represents an object, and the numbered circles signify all object attributes. Some of these attributes are measurable, and some are unquantifiable. Also, some are common attributes, while some attributes are unique. The shaded attributes 1, 5, 7, 10, 14, and 15 are QIs that can be used to differentiate an object from others of a similar kind.

This paper classifies the attributes into the following three segments:
\begin{enumerate}
    \item \textbf{Static attributes: } Every object possesses the kind of properties that are mandatory, have to remain unchanged and constitute towards its uniqueness. These properties are crucial for the overall identity of the object. Meaning, any change in these attributes would change the complete identity of the object. For instance; in case of of a famous tourist sight its location is elementary. Moving it to somewhere else, it may destroy its identity.
    
    \item \textbf{Dynamic attributes: } There are some properties in every object that may change over time in a predictable extent. For example, the color of a car may bleach over time and falls under this category. Change in color give new appearance to the object but the object will still exhibit the same behavior during interaction. Therefore, it is safe to say that the overall identity of the object did not change. It is still the same car in the example above.
    
    \item \textbf{Volatile attributes: } With usage, the object witnesses changes in volatile properties. These properties are occurring most frequently and are least affecting to the identity. Signs of wear and tear, minor scratches on surface and interior parts are common examples of volatile attributes. They are only important as an overall set, defined by certain criteria such as its majority.
\end{enumerate}

Using the attributes collectively, it is possible to create an identifier that reflects the identity of an object at all times. Although conventionally identifiers are linked with an object but it did not represent the identity of that object. But recently, there have been attempts towards attribute based identification schemes. For e.g; use of parts of the content to uniquely identify a document, or images of objects to extract possible attributes and use it for identification. Some of the common  \cite{Bkheet2021ARO.identification_methods} conventional and modern identification methods are summarised in the following subsection. 

\subsection{Identification methods}
\begin{enumerate}
    \item \textbf{Conventional identification methods}
    
    \begin{enumerate}
          \item Radio Frequency Identification (RFID): First developed in 1945, RFID uses radio waves to identify objects. The RFID consists of a tag, reader, antenna, server, access controller and software. The RFID tag can be attached to any object as an identifier. RFID provides a mechanism to detect, track, and control the object electronically. The identifier for an RFID is the electromagnetic/electrostatic coupling in the radio frequency portion of the electromagnetic spectrum. The object that wants to be identified sends uniquely identifiable electromagnetic signals using the tag. The signals are detected by the reader on the verifier's side.
        
        \item Barcode: Barcode is another prevalent method that can be easily found on commercial products. A barcode is an optically machine-readable label printed or attached to the objects. It is a representation of unique alphanumerical or numerical code. Straight lines of different widths on a contrasting background depict these codes. Figure \ref{fig:barcode} shows barcode signatures of alphabets and numerals. Any identifier can be translated to a barcode following the standard translation. 
        
        \begin{figure}[h]
            \centering            \includegraphics[scale=0.2]{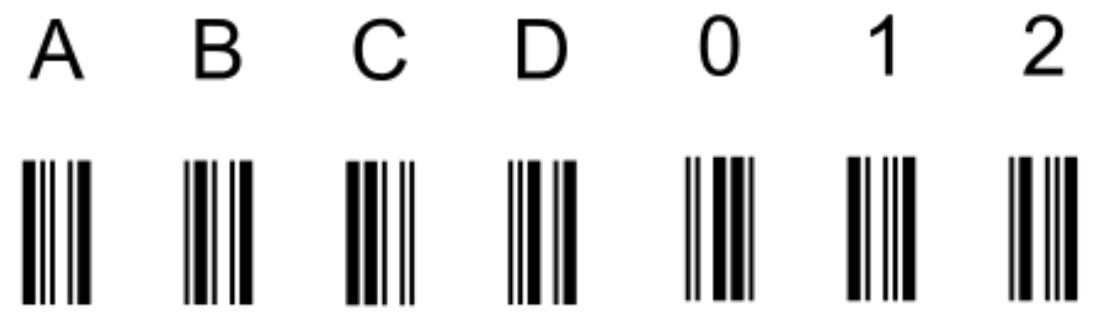}
            \caption{Barcode translation of English alphabets and numbers}
            \label{fig:barcode}
        \end{figure}
        
        \item QR-code: It can be seen as an advanced version of the barcode. It is a square-shaped optically-readable representation of complex data structures like JSON. Where barcodes could only represent letters, digits and a few special characters, QR codes can hold a comparatively large amount of information like website URL/URI or cryptographic keys.
       
        \item IPv4/v6 addressing: IPv4 is an old version of the currently used internet protocol (IP) --- IPv6. An IP address is used to identify and locate an object on a network. IP works as a network identifier for the objects. IPv6 is a 128-bit long address where the starting 64 bits are network prefixes and the rest 64 bits are the interface's identifier.
    \end{enumerate}
    
    \item \textbf{Modern identification methods}
    \begin{enumerate}
        \item Electronic product code (EPC): EPC is one of the most prevalent object naming systems. It is a universal identifier for uniquely identifying a vast number of objects in this world. For this reason, the EPC code should be reasonably large to allow numerous objects. EPC constitutes hierarchical sections. EPC is 96-bits long, out of which the first 8-bits hint at EPC type. Further, three sections consist of an EPC manager, an EPC object class and a serial number. Two additional sections are dedicated as an identifier of the object. Following is a general representation of a typical EPC:
        {\verb|type:manager:class:serial:id1.id2|}
        
        \item Fingerprint-of-things: The identifier created using information about the object comes under the fingerprint of things. It creates a unique signature called fingerprints based on information like software and hardware components of the object. Other types of attributes are also used, like the image of surface-pattern \cite{Rui2017IndividualRB.surface}, sequence of packet-transfers during the start-up of a device \cite{Aluthge2018IoTDFI.packet}, network traffic analysis \cite{Bezawada2018IoTSenseBF.networkTraffic}. Further on, integrating different attributes and with the help of machine learning, it is argued \cite{Vaidya2020IoTIDAN.iotID} that a device-specific identifier can be created. However, the concept is still novel and going through changes.
        
        \item Machine Learning methods: In recent attempts to object identification using machine learning, the data scientists study the network communication pattern of various IoT devices. The study \cite{Meidan2017DetectionOU.IoTidentification} applies supervised classification machine learning algorithms to the network traffic data of individual objects to classify the type of device. Whereas in the study \cite{Meidan2017ProfilIoTAM.IoTandML} the experiment was performed on nine objects. The data collected from these devices were used to train and evaluate the algorithm in two stages. In the first stage, the classifier distinguishes IoT devices from non-IoT devices. The devices are linked to a specific IoT class in the next stage. 
        
    \end{enumerate}
\end{enumerate}

The conventional identification methods primarily focus on devising an identifier without considering the properties of an object. There is no association with the object's identity. On the other hand, modern identification methods use the object's attributes to derive the identifiers. The current research extends this effort by proposing an attribute encoding technique. The attributes can be collectively encoded to produce an identifier that fulfils the requirements of an identifier.

\subsection{Requirements of an identifier}
\label{subsec:requirements}
An identifier must consider several features during its design. The NIST \footnote{National Institute of Standards and Technology} Computer Security Handbook \cite{NIST.CIAtriad} lists information security objectives. These objectives are often known as the CIA triad and further extended to CIAAAR. The CIAAAR is an abbreviation of six security objectives: confidentiality, integrity, availability, authenticity, accountability and non-repudiation. The requirements addressed by the identifier in terms of CIAAAR can be summarised as follows:
\begin{enumerate}
    \item \textbf{Confidentiality: } Authorised restriction over a piece of information with an intent to preserve personal privacy.
    \begin{itemize}
        \item [Req 1a] \textbf{Anonymity: }The identifier should never allow unauthorised disclosure of the owner's identity.
        \item [Req 1b] \textbf{Unlinkability: }As desfined by ISO in privacy framework~\footnote{\url{https://www.iso.org/standard/45123.html}, Accessed: 20 May 2022} inability of connecting two identifiers to a common QI. It should not be the case that identifiers of two objects have common bits for the same attributes. It ensures that nobody can determine the attributes by looking at their identifiers. Therefore, unlinkability is an important feature for maintaining confidentiality.
    \end{itemize}
    
    \item \textbf{Integrity: }Preservation of information against unauthorised modification and deletion.
    \begin{itemize}
        \item [Req 2a] \textbf{Immutability: }The identifier should remain unaltered unless changed via an authorised process.
    \end{itemize}
    
    \item \textbf{Availability: } Reliable and timely access to (use) information.
    \begin{itemize}
        \item [Req 3a]\textbf{Verifiability: }The identifier should be verifiable by a third party at all times.
    \end{itemize}
    
    \item \textbf{Authenticity: }The ability to be trusted so that there is confidence in the validity of the information. The genuineness of the information should be verifiable.
    \begin{itemize}
        \item [Req 4a]\textbf{Flexible: }The identifier must represent the object's identity. Therefore, the identifier should reflect which attributes of the object have changed by changing itself partially or fully. So, the identifier must always be the same, no matter where it is calculated. 
        \begin{itemize}
            \item [Req 4a1]\textbf{Dynamic part: }The identifier part responsible for representing dynamic attributes should change without affecting the rest of the identifier. For example, "4z3agf2h5" is an identifier. "4z3", "agf", "2h5" signify three attributes of an object. If the attribute represented by "agf" is dynamic, its modification should make the identifier like "4ztk9nh5". It should be noted that the initial and final bits of strings are the same as in the previous version of it. The section \ref{sec:model} discusses this concept in detail.
            \item [Req 4a2]\textbf{Static part: }The part of the identifier representing static attributes should alter the whole identifier if changed.
        \end{itemize}
    \end{itemize}
    
    \item \textbf{Accountability: }The action of an entity should be traced uniquely to that entity.
    \begin{itemize}
        \item [Req 5a] \textbf{Uniqueness: } Each identifier must be linked with a single object. In other words, Each object must possess a unique identifier.
    \end{itemize}
    
    In addition to CIAAA, experts link another objective to the list for covering additional aspects of data security:
    \item \textbf{Non-Repudiation: }An entity should not be able to deny responsibility for its action.
    \begin{itemize}
        \item [Req 6a]\textbf{Investigability: }The identifier must betray the shifts in the object properties compared to its previous version. It ensures that the change in an object’s identity never goes unrecognised.
    \end{itemize}
    
\end{enumerate}

\section{Identifier calculation}
\label{sec:identifier_creation}

This section discusses popular hashing algorithms and different hash-based identification methods.

In the case of document fingerprinting \cite{Schleimer2003WinnowingLA.dochash}, the document's paragraphs need to be encoded. Also, while storing images in the database \cite{Chum2008NearDI.imagehash}, they are textually encoded to ensure non-redundancy and easier management. Similarly, the git version control system uses hashes \cite{Chacon2014ProG.githash} to name its contents like files, directories and revisions. 
\subsection{Cryptographic Hashing algorithms}
     Cryptography comprises several mathematical and computational processes to design an encryption and decryption technology to provide data security and authentication \cite{Meyer1989CryptographyaSO}. Hashing is a family of cryptographic algorithms that offer only one-way encryption, meaning the hash strings cannot be decrypted. A hashing algorithm, generally, takes an arbitrary length of data, applies a formula and produces a fixed length output --- called a hash value. The quality of a hash is primarily observed by the probability of collision of two hash values. Collision is when the same hash values are calculated for more than one input. If the collisions occur frequently, then an adversary has the possibility to reverse engineer the input using brute force. Hence, failing the purpose of one-way encryption.
    Some of the most common hashing algorithms are: MD5, SHA2 and NTLM  \cite{ROUNTREE201129.hashing}:
    
    \begin{itemize}
        \item MD5: Message digest (MD) is a family of the algorithm. MD5 is the fifth version of the family. MD5 generates 128 bits long output. After witnessing wide use, it phased out due to higher collision rates. 
        \item SHA2: SHA2 is a suite of algorithms including SHA-224, SHA-256, SHA-512. The suffixed numbers represent the length of hash values. SHA256 is widely used in many sectors like banking, blockchain, and version control. SHA256 algorithm is initialised with the first 32 bits of the fractional part of the square roots of the first eight natural prime numbers. Following the initialisation, there are mathematical and computational steps that make use of every bit of the input. In addition, the same steps are applied 64 times, ensuring that even a slight input change alters the hash value.
        \item NTLM: Successor of Microsoft LANMANv2 hashing algorithm, NT Lan manager algorithm (NTLM) is used for password hashing during authentication in windows systems. The second version of NTLM (NTLMv2) uses the HMAC-MD5 algorithm for hashing.
    \end{itemize}
    
\subsection{Hash-based identification methods}
\begin{itemize}

    \item Document fingerprinting: A document fingerprint \cite{Manber1994FindingSF.docfingerprint} is an encoded alphanumeric string representing part of a document. Typically, a substring is selected from the text and hashing function is applied to it. Similarly, a hash of several substrings from the document is stored in an index. Collectively, all hashes together form a unique identity. Hash sets of two documents can be compared to detect duplicity or plagiarism. 
   \item Robust hashing: A hash is extremely sensitive to the input. Even the slightest modification in the input produces a different hash. Robust hashing is a modification over conventional hashes. Robust hashes produce the same result if the inputs are changed only slightly. The robust hashes were first developed for images \cite{venkatesan2000robust}. In an attempt to optimise the storage in databases, robust hashes helped reduce redundancy from storing visually similar images.
    
    Together with optical character recognition (OCR), Robust hashing finds its use in texts \cite{villan2007tamper.robusthashing.text}. In OCR based approach, the text is optically read and hashed. If there is any minor modification in the text, it is ignored and the hash remains unchanged.
    
    In other research, \cite{steinebach2013robust.text}, robust text-based hashing is used for document fingerprinting and plagiarism checks. Some part of the document's text is hashed and used as a reference. Later, hamming distance \cite{Norouzi2012HammingDM} of two hashes is compared and if it is less than the threshold, that part of the document is a successful match.
    
%    \item Git: The prominent version controlling system also used hashing to identify the files and their versions \cite{Chacon2014ProG.githash}. However, it should be noted that all version control systems follow different strategies. It is common to refer to different file versions using sequential numbers. The advantages of using hashes are: 1. Integrity checks as bit flip or corrupted contents are easily detectable; 2. Object lookups are faster.
%    Additionally, the git protocol uses a hash to address the stored contents reliably. Initially, SHA1 was used as the hashing algorithm, but later, it shifted to SHA256.
    
%    \subsubsection{Other applications: }Hashing can be used together with other encryption methods to offer some widely used applications \cite{ROUNTREE201129.hashing}. Digital rights management (DRM) systems use the hash to ensure authorised content access. Pretty good privacy (PGP) uses hash for the confidentiality of emails. The digital signature is a user's digital identity, proven with the ownership of a private key.
\end{itemize}
      
    \begin{figure*}
        \centering
        \includegraphics[scale=0.25]{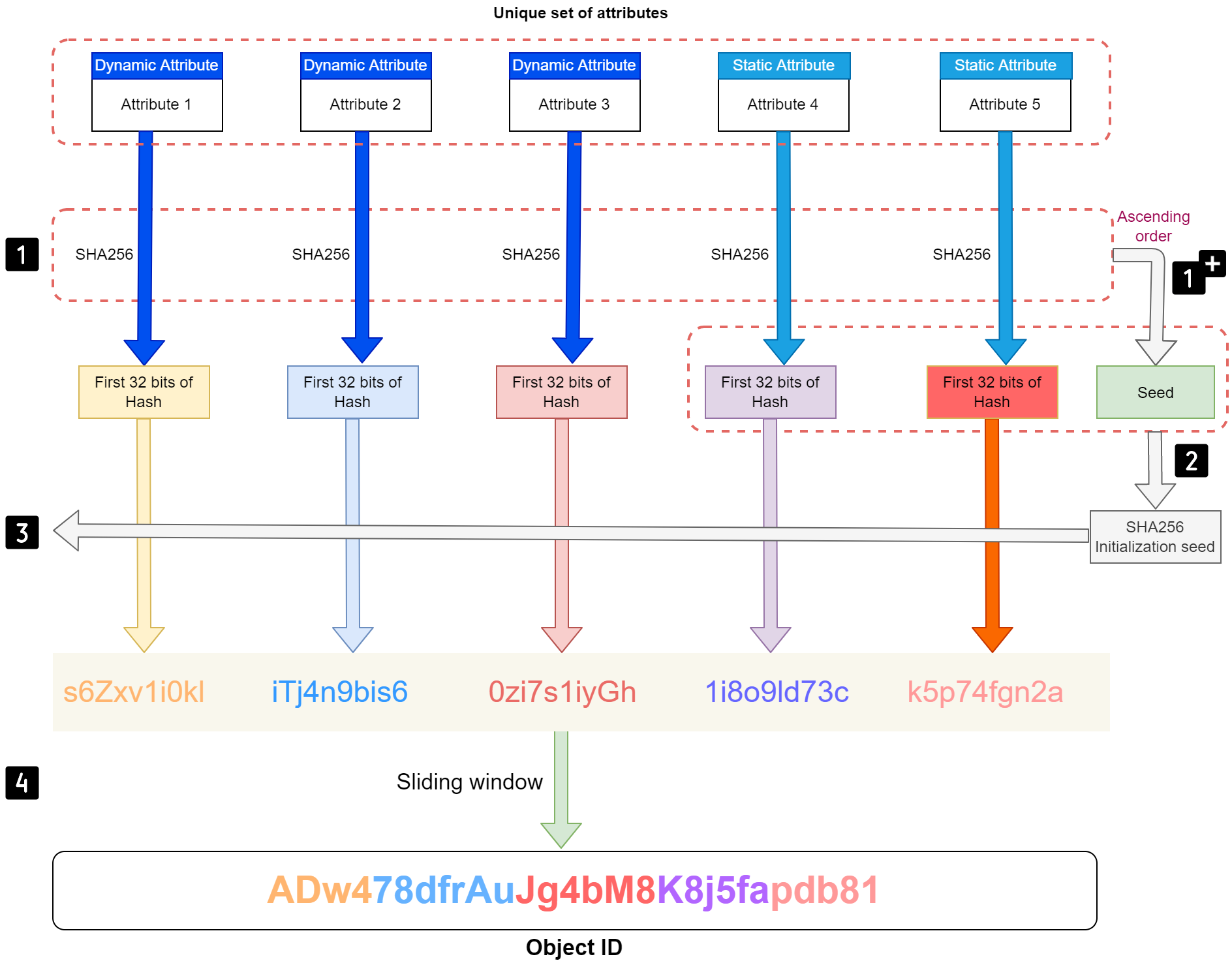}
        \caption{Steps to calculate attribute based Object identifier}
        \label{fig:model_algorithm}
    \end{figure*}  
      
\section{Algorithmic model}
\label{sec:model}

\begin{figure}
    \centering
    \includegraphics[scale=0.6]{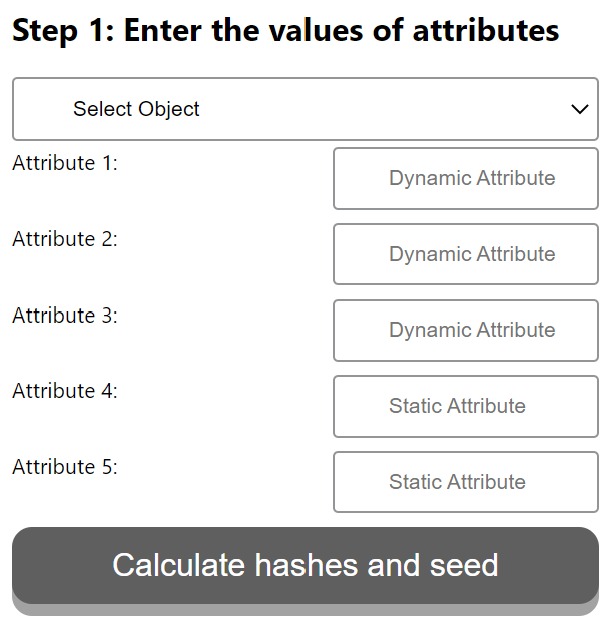}
    \caption{Interface to feed the values of attributes}
    \label{fig:objectID_UI}
\end{figure}

This section describes the proposed algorithm for producing attribute-based identifiers in the form of advanced robust hashes. Figure \ref{fig:model_algorithm} depicts the sequential process to generate an object identifier. The object is assumed to have five QI. The input can be provided in any textual form. In the current implementation, the attributes can be entered into an interface by the user, as shown in figure \ref{fig:objectID_UI}. 

\begin{table*}
\centering
\begin{tabular}{|l|l|l|l|l|l|l|} 
\hline
          & Attribute 1           & Attribute 2     & Attribute 3 & Attribute 4       & Attribute 5     & Seed            \\ 
\hline
Attribute & Operating Temperature & Working sensors & IP address  & Mac address       & Type of sensors & -               \\ 
\hline
Value     & 28.60                 & $4$             & 0.0.0.0:00  & e5:84:e6:2f:33:61 & infrared        & -               \\ 
\hline
Step 1    & e9162517              & 4b227777        & 66973d16    & d99d68359         & bd0f2196        & -               \\ 
\hline
Step 1+   & -                     & -               & -           & -                 & -               & b07de73m        \\ 
\hline
Step 2    & -                     & -               & -           & \multicolumn{3}{l|}{e15b2bcb110d0b2ef3e52274a91527e}  \\ 
\hline
Step 3    & s6Zxv1i0kl            & iTj49bis6       & 0zi7s1iyGh  & 1i8o9Id73c        & k5p74fgn2a      & -                \\ 
\hline
Step 4    & \multicolumn{5}{l|}{ADw478dfrAuJg4bM8K8j5fapdb81}                                           & -                 \\ 
\hline

\end{tabular}

\caption{Example: Calculating Object ID from its attributes for the first time \label{tab:objectIDgeneration1}}
\end{table*}

\begin{table*}
\centering
\begin{tabular}{|l|l|l|l|l|l|l|} 
\hline
& Attribute 1 & \textbf{Attribute 2} & Attribute 3 & Attribute 4  & Attribute 5 & Seed \\ 
\hline

Attribute & Operating Temperature & \textbf{Working sensors} & IP address  & Mac address       & Type of sensors & -               \\ 

\hline
Value  & 28.60   & \textbf{3}  & 0.0.0.0:00  & e5:84:e6:2f:33:61 & infrared  & -      \\ 

\hline
Step 1 & e9162517  & \textbf{4e074085} & 66973d16  & d99d68359  & bd0f2196   & -   \\ 

\hline
Step 1+   & -  & -      & -      & -      & -           & b07de73m        \\ 

\hline
Step 2    & -            & -       & -           & 

\multicolumn{3}{l|}{e15b2bcb110d0b2ef3e52274a91527e}  \\ 
\hline
Step 3    & s6Zxv1i0kl            & \textbf{ef28766d}       & 0zi7s1iyGh  & 1i8o9Id73c        & k5p74fgn2a      & -                \\ 
\hline
Step 4    & \multicolumn{5}{l|}{ADw47\textbf{b890e88f}4bM8K8j5fapdb81}                                           & -                 \\ 
\hline

\end{tabular}

\caption{Example: Calculating Object ID from its attributes for the first time \label{tab:objectIDgeneration2}}
\end{table*}

\subsection{Attribute-based object ID model}
In figure \ref{fig:model_algorithm}, the dynamic attributes are represented in dark blue coloured boxes and arrows, whereas static attributes are shown in cyan colour. The calculation majorly involves four steps. The explanation of the processing of the algorithm is stated as under:
\begin{enumerate}
    \item [ S $1$ .] \textbf{Encryption} Initially, each attribute is hashed separately using SHA256. For simplicity and ease of further processing, the hash values are trimmed to take only the first 32 bits to the next steps. 
    
    \item [ S $1^+$.] \textbf{Seeding} This step only occurs once. While generating the object identifier for the first time, individual hashes of all attributes are used to calculate the seed. This seed is stored separately on the cloud or blockchain.
    
    \item [ S $2$ .] \textbf{Initialisation} First 32 bits of the hashes of static attributes, together with seed, are used to initialise the SHA256 algorithm. These bits are split into eight equal-length bits and used as seeds for the SHA256 hashing algorithm. 
    
    \item [ S $3$ .] \textbf{Diffused encryption} Now, the modified SHA256 algorithm from step 2 is used to calculate the hash of step 1’s outcome individually. Again, the first 32 bits of each hash value are concatenated together.
    % the hash algorithm is diffused after replacing the new seeds - hence diffused encryption https://en.wikipedia.org/wiki/Confusion_and_diffusion
    
    \item [ S $4$ .] \textbf{Fuzzification} Sliding window operation of 16-bit size is applied over the whole concatenated hashes from step $3$. In this operation, the first 16 bits are taken. At first, the ASCII value of each bit is summed together. For modulus operation, the sum is divided by the value 123 because the ASCII value for the last alphabet ends at 122. After the modulus is applied, there is a criterion-based calculation stated below:
    \begin{itemize}
        \item If the sum is less than 48, then 48 is added to the sum.
        \item If the sum is in the range [58,64], then 7 is added to the sum.
        \item If the sum is in the range [91,96], then 8 is added to the sum.
    \end{itemize}
    These criteria ensure that the sum always stays in the alphanumeric zone in ASCII representation.  
\end{enumerate}

\subsection{Object ID calculation: Example}
The table \ref{tab:objectIDgeneration1} demonstrates the generation of object identifiers based on the proposed model. The object considered for this example is a 3D vision sensor. The first three attributes, i.e., Attribute 1, Attribute 2 and Attribute 3 are considered dynamic attributes, whereas Attribute 4 and Attribute 5 represent static attributes. For demonstration, the uniquely identifiable properties and their values have been assumed in the subsequent rows. 

In step 1, there are the first 32 bits of the hashes of the attribute values. 
Since this is the first time calculating the identifier, step $1^+$ must create a seed using all the attributes' full hashes.
In step 2, the hash of attributes 4 and 5 --- which are static attributes and the seed are concatenated and hashed using SHA256 to calculate the seed for the next step.
Step 3 uses the new seed to modify the SHA256 algorithm. All the hashes from step 1 are hashed separately using the modified SHA256 algorithm.
In step 4, all hashes from step 3 are concatenated. Now, 16 bits sliding window covers the whole hash by doing the following calculations:
\begin{itemize}
    \item Sum of ASCII values of s,6,Z,x = 379
    \item Mod by 123 = 379\%123 = 10
    \item Since the value is less than 48: new value = 10 + 48 = 58
    \item Since the new value falls in the range [58,65], new value = 58 + 7 = 65
    \item 65 is the ASCII value for A.
    \item Similarly, the sliding window moves forward, calculates "6Zxv", and gets the outcome --- D. 
\end{itemize}

Additionally, the whole identifier is calculated as stated in step 4 in the table.

Now, if Attribute 2 of the object alters from '4' working sensors to '3', subsequently, the hash corresponding to Attribute 2 updates following the proposed algorithm. It is evident in the new identifier presented in table \ref{tab:objectIDgeneration2} that part of the object identity has changed. It should be noted that only part of the identifier has changed, signalling that the attribute was dynamic.

Lastly, in case of Attribute 4 or 5 is altered, the whole identifier changes because these are the static attributes.
    
\section{Analysis}

This section briefly analyses the FaR hash algorithm with standard hashing algorithms, SaR hash and robust hashing techniques. It can be discussed considering two possible scenarios based on the type of input modification: dynamic or static. In the first case, a dynamic attribute has changed. As a result, standard hashing algorithms like SHA256 will produce an identifier that exhibits the Hamming distance of 1. In the case of robust hashing, the Hamming distance is potentially 0, as robust hashing is designed to ignore small changes in input. In SaR Hash approach, the Hamming distance will be more than 0 but still significantly less than 1. Lastly, FaR hash will show Hamming distance slightly more than that of in the case of SaR. The difference between SaR and FaR hashes Hamming distances is attributed to the fuzzified bits around the particular dynamic attributes neighbourhood. 

The higher Hamming distance allows extra room for the diffusion of a QI. These extra bits should allow more hashes to be calculated before the first hash collision. Mathematical and experimental arguments for this statement are part of future studies.

\section{Discussion}

Each step of the FaR hash development attempts to fulfil one or more of the CIAAA+R requirements discussed in the section \ref{subsec:requirements}. The table \ref{tab:step_requirements} summarises the significance of each step of the proposed algorithm towards the requirements.

Anonymity \textit{(Req 1a)} is ensured by using hashes during step \textit{S 1}. By design, hashes unidirectionally encode plain text. An attribute is hashed to hide the information about that object. Although the usage of hash is not just limited to anonymity, it also facilitates further calculation of the proposed model. The next step \textit{Req 1+} fulfills two requirements: unlinkability \textit{(Req 1b)} and immutability \textit{(Req 2a)}. As assumed initially, each object has at least one uncommon QI and thus unique seeds. Using the seed further in the algorithm ensures that encryption of the same QI of two distinct objects is different. For example, a QI, viz. "type of sensors," is "infrared" for multiple objects. Since their seeds are unique, the part of the object identifier signifying this QI will be different. Thus, it is impossible to guess an attribute by comparing multiple object identifiers and accomplishing unlinkability.
On the other hand, the immutability is attempted by using \textit{Step S1+} with \textit{Step S2}. The seed is calculated using all the hashes of the attributes in \textit{Step S1+} and then, using the seed and static attributes for initialising the hash algorithm in \textit{Step S2} --- it modifies the hashing algorithm with deterministically random seeds. It has not been experimentally analysed, but the modified hashing algorithm should logically exhibit diffusion in the outputs. The diffusion in the output bits produces unique strings. Additionally, the produced strings will be distinct from its counterpart resulting from a standard hashing algorithm with no change in the seeds. Therefore, the seed can also be used as a secret for the whole calculation. During the idea's further development, the seed can be stored on the blockchain. This way, it will remain publicly accessible and hence third-party verifiable. While immutable from unauthorised manipulation during the life-cycle of the object.

Subsequently, the \textit{Step S2} and \textit{Step S3} together address the requirement \textit{Req 4a2} i.e., (flexible: static part). The modified hash algorithm is applied over the hashes of all attributes individually. The modification of hashing algorithm includes static hashes along with the initial seed. Consequently, the hashing algorithm will be initialised with the updated seeds if any static attribute changes. Thus, producing different results for all attributes compared to the previous version(s).

On the other hand, Req 4a1 (flexible: dynamic part) is achieved by hashing the dynamic QIs in isolation during \textit{Step 3}. The isolated hashing of each QI prevents the effects of a changing QI on the other parts of the identifier. \textit{Step 3} and \textit{Step 4} together bring investigability (Req 6a) by indicating the changed QI of the object. The isolated hashing in step 3 and fuzzification brought by the sliding window computation are the main components of this algorithm. As discussed earlier, the step \textit{Step 2} and \textit{Step 3} distinguishes static and dynamic QIs. Now, in \textit{Step 4}, fuzzification diffuses the change of a dynamic QI to the neighbouring bits. The output coming out of the \textit{Step 4} is capable of changing partially (in case of dynamic QI) or fully (in case of static QI).

After the identifier has been published on a public blockchain, all the object versions will remain visible. Additionally, the seeds can be stored in an encrypted format, which the user's private key can decrypt. Together with the algorithm and seed, the object identifier will remain third-party verifiable (Req 3a). 

As each object has a unique set of attributes, and standard hashes output almost unique hashes, the algorithm fulfils Req 5a (uniqueness). In addition to the logical proposition, the requirements must also be established experimentally and, where applicable, mathematically, which will be part of future research.

\begin{table}
\centering
\begin{tabular}{|l|l|} 
\hline

\textbf{Steps} & \textbf{Requirements} \\ 
\hline

S 1 & Req1a   \\ 
\hline

S 1+ & Req1b, Req2a   \\ 
\hline

S 2 & Req2a, Req4a2   \\ 
\hline

S 3 & Req4a1, Req4a2, Req6a    \\ 
\hline

S 4 & Req6a   \\ 
\hline

% req 1a  1
% req 1b 1+
% req 2a 1+, 2
% req 3a 
% req 4a1 3
% req 4a2 2, 3
% re1 5a 
% re 6a 3, 4

\end{tabular}

\caption{Significance of steps in the algorithm model towards fulfillment of requirements
\label{tab:step_requirements}}
\end{table}

\section{Future Work and Conclusion}

The paper introduced a novel concept called FaR hash that demonstrates a significantly different behaviour than a standard hash. On the one hand, the standard hashes present an entirely new set of bits on slight modifications in the input; on the other hand, the FaR hash is capable of selective updates in the output bits of the hashes. That is, the hash changes based on the type and position of the input. This property of the FaR hash aligns perfectly with the requirements of a CIAAA+R compliant object identifier. 

The paper extends the idea that the object identifier directly represents the object’s identity. An identity is the state of being oneself and can be defined by the combination of its attributes or QI. These QIs are classified into three types --- static, dynamic and volatile. Static QIs are the required QIs. Dynamic QIs are comparatively less important attributes and may vary within an predictable extent. Lastly, volatile QIs may represent short-term or permanently changing QIs which are only important as a total set. In the proposed algorithm of the FaR hash, the textual values of the QIs are the inputs. If the static QIs are modified, the object identifier changes entirely. While if a dynamic QI changes, only the corresponding part of the identifier changes.

Although currently, the number of inputs is pre-defined during the implementation, being flexible with the inputs can enhance the application areas of FaR hash. Another aim of future work will be to examine the effect of modifying the standard hash algorithm with new seeds during the \textit{Step 2} in the algorithm (as shown in figure \ref{fig:model_algorithm}. This paper introduces the idea of using FaR hash as the object identifier, but it should be mathematically and experimentally established in the future.

\bibliographystyle{ACM-Reference-Format}

\end{document}